\documentclass[prl,superscriptaddress,twocolumn,showpacs,preprintnumbers,amsmath,amssymb]{revtex4}

\usepackage{graphicx}

\begin{document}


\title{Production of Ultracold, Polar RbCs$^{*}$ Molecules via Photoassociation}

\author{Andrew J. Kerman}
\author{Jeremy M. Sage}
\author{Sunil Sainis}
\affiliation{Department of Physics, Yale University, New Haven, CT 06520, USA}
\author{Thomas Bergeman}
\affiliation{Department of Physics and Astronomy, SUNY, Stony Brook, NY 11794-3800, USA}
\author{David DeMille}
\affiliation{Department of Physics, Yale University, New Haven, CT 06520, USA}

\date{\today}

\begin{abstract}
We have produced ultracold, polar RbCs$^*$ molecules via
photoassociation in a laser-cooled mixture of Rb and Cs atoms.
Using a model of the RbCs$^*$ molecular interaction which
reproduces the observed rovibrational structure, we infer decay
rates in our experiments into deeply bound X$^1\Sigma^+$ ground
state RbCs vibrational levels as high as 5$\times 10^5$ s$^{-1}$
per level. Population in such deeply bound levels could be
efficiently transferred to the vibrational ground state using a
single stimulated Raman transition, opening the possibility to
create large samples of stable, ultracold polar molecules.
\end{abstract}



\maketitle

Ultracold polar molecules, due to their strong, long-range,
anisotropic dipole-dipole interactions, may provide access to
qualitatively new regimes previously inaccessible to ultracold
atomic and molecular systems. For example, they might be used as
the qubits of a scalable quantum computer \cite{Qcomp}. New types
of highly-correlated many-body quantum states could become
accessible such as BCS-like superfluids \cite{BCS}, supersolid and
checkerboard states \cite{dipolar}, or ``electronic" liquid crystal phases
\cite{LCD}. Ultracold chemical reactions between polar
molecules have been discussed \cite{coldchem}, and might be
controlled using electric fields \cite{linking}. Finally, the
sensitivity of current molecule-based searches for violations of
fundamental symmetries \cite{EDM} might be increased to
unprecedented levels.

Cold, trapped polar molecules have so far only been produced using
either buffer-gas cooling \cite{buffer} or Stark-slowing
\cite{stark}, at temperatures of $\sim$10-100 mK \cite{buffer,
stark}. This is much higher than the $\sim$1-100  $\mu$K
accessible with atoms, and attempts to bridge this gap with
evaporative cooling may run afoul of predicted molecular Feshbach
resonances \cite{molfesh} or inelastic losses \cite{molinel}.

Another approach is to extend well-known techniques for producing
ultracold (non-polar) homonuclear diatomic molecules in binary
collisions of ultracold atoms, either through photoassociation
\cite{Cs2,K2,Rb2,heinzen,Na2}, or Feshbach resonance
\cite{chu,feshBEC}. In these methods, the translational and
rotational temperatures of the molecules are limited only by the
initial atomic sample, possibly providing access all the way to the
quantum-degenerate regime \cite{heinzen,feshBEC}. An important
limitation, however, is that the molecules are typically formed in
weakly bound vibrational levels near dissociation, which may have vanishing electric dipole 
moments \cite{jul03}, and are unstable with respect
to inelastic collisions \cite{heinzen,molfesh,molinel}; therefore,
a method for transferring them to the vibrational ground state is
desirable \cite{K2}.

Several authors have discussed the extension of these methods to
the formation of (heteronuclear) polar molecules in collisions 
between different atomic species \cite{wangstwa,ions,Li6Li7,dipsup}. 
In recent experiments NaCs$^+$ and RbCs$^+$ ions formed in the presence of
near-resonant light have indeed been observed in small numbers
\cite{ions}; however, these observations did not permit an
analysis of their formation mechanism, nor demonstrate a method
for producing neutral, ultracold polar molecules.

In this Letter, we describe the production of electronically
excited, polar RbCs$^*$ molecules via photoassociation in an
ultracold ($T\sim 100 \mu$K) mixture of $^{85}$Rb and $^{133}$Cs
atoms. We have observed their electronic, vibrational, rotational,
and hyperfine structure, as well as the large DC Stark effect
characteristic of a polar molecule. Analysis of our data allows us
to infer spontaneous decay rates to deeply bound vibrational
levels of the RbCs X$^1\Sigma^+$ electronic ground state of up
to $\sim5\times 10^5$ s$^{-1}$ per level. Our calculations show
that molecules in such levels could be transferred to the
vibrational ground state of RbCs with a single, stimulated Raman
transition.

\begin{figure}
\includegraphics[width=3.375in]{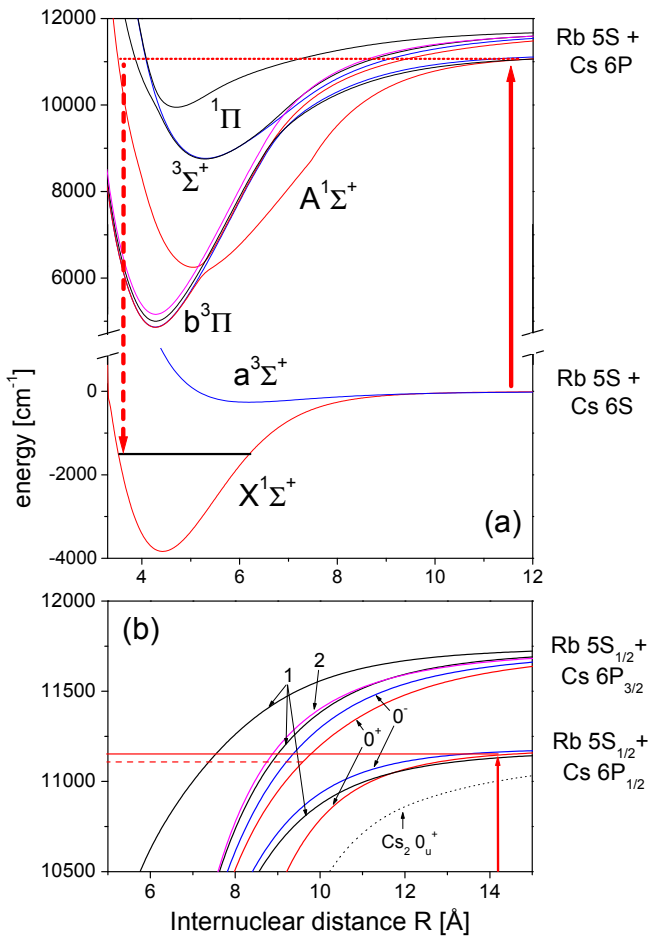}
\caption{(Color online) Schematic of the photoassociation process. (a) $^{85}$Rb +$^{133}$Cs atom pairs are excited during a collision (upward arrow) to molecular levels (horizontal line) below the Rb 5S$_{1/2}$ + Cs 6P$_{1/2}$ atomic asymptote. They can then decay (dashed, downward-pointing arrow) to ground state molecular levels. (b) Detailed view of the potentials, labelled by their Hund's case (c) quantum number $\Omega=0^+,0^-,1,2$. The horizontal lines indicate vibrational levels accessed in this work. The dotted curve shows the potential for the $0_u^+$ state of Cs$_2$, to illustrate its longer range character for weakly bound levels.}
\label{fig1}
\end{figure}

Photoassociation (PA), illustrated in Fig. \ref{fig1}, occurs when
two colliding ground state atoms absorb a photon and are promoted
to a weakly bound, electronically excited molecular level
\cite{molsum,bohnPA}. As indicated in Fig. \ref{fig1}(b), the
levels accessed in heteronuclear PA are of much shorter range than
their homonuclear counterparts; this arises from the fact that at
long range, in their first excited state, two identical atoms
interact via the resonant-dipole interaction (with potential
$V(R)\propto R^{-3}$), whereas two atoms of different species
interact only via much shorter-ranged van der Waals forces
($V(R)\propto R^{-6}$) \cite{wangstwa,marsad}. In the latter case, for a given excited-state binding energy,
the Franck-Condon factor (FCF) characterizing the overlap between the initial free-atom ground state
and the excited molecular bound state is significantly smaller, requiring a higher PA
intensity. However, the FCF for decay to deeply bound vibrational
levels of the ground X$^1\Sigma^+$ state is also larger, as we
discuss below.

Our observations are made in a dual-species magneto-optical trap
(MOT) \cite{epaps}. We excite colliding atoms into RbCs$^*$
rovibrational levels at a variable detuning $\Delta$ below the
lowest lying excited asymptote, correlating to Rb 5S$_{1/2}$+Cs
6P$_{1/2}$ (this is the most favorable choice for our purposes
since these levels do not predissociate \cite{prediss}). After a
pair is excited, it decays either to a ground-state RbCs molecule,
or back to energetic free Rb and Cs atoms which typically escape
from the two traps. The signature of RbCs$^*$ formation is then a
resonant steady-state depletion of both the Rb and Cs traps
induced by the PA laser. For optimal sensitivity, we maximize the
PA-induced loss rate (proportional to the product of the Rb and Cs
densities integrated over the PA beam profile) while minimizing
other intrinsic losses that compete with it to determine the
steady-state trap populations. For this purpose, we use forced
dark-spot MOTs \cite{darkspot,radcol} to increase the Rb and Cs densities
by factors of 9 and 4 (relative to ``bright" MOTs), respectively,
while reducing their intrinsic losses due to light-assisted
inelastic collisions \cite{radcol}. These losses are further
reduced in the Rb trap by significantly decreasing the trapping
laser intensity.

The peak density $n$, atom number $N$, and spatial overlap of the
two atomic clouds were optimized using two-color absorptive
imaging from two orthogonal directions. Typical values were:
$N_{Rb} = 4\times 10^8$, $n_{Rb} = 7\times 10^{11}$ cm$^{-3}$, and
$N_{Cs} = 3\times 10^8$, $n_{Cs}=3\times 10^{11}$ cm$^{-3}$.
Independent measurements of $N_{Rb,Cs}$ using resonance
fluorescence confirmed these values to within $\sim$30\%. Temperatures of $T_{Rb}=55\;\mu$K and
$T_{Cs}=140\;\mu$K were measured using time-of-flight absorption imaging \cite{csint}.

The Ti:sapphire PA laser produced 600 mW of power around 895 nm;
its frequency was monitored using both a wavemeter and an optical
spectrum analyzer, providing absolute (relative) accuracy of 150
MHz (5 MHz). To increase the PA intensity, the laser light was
modematched into a build-up cavity (finesse $\sim 60$) placed
around the vacuum chamber; its $e^{-2}$ mode radius at the atom
traps of 380 $\mu$m (the typical size of the MOT clouds was 750
$\mu$m) resulted in PA intensities up to $4\times 10^7$ W/m$^2$
\cite{epaps}.

\begin{figure}
\includegraphics[width=3.375in]{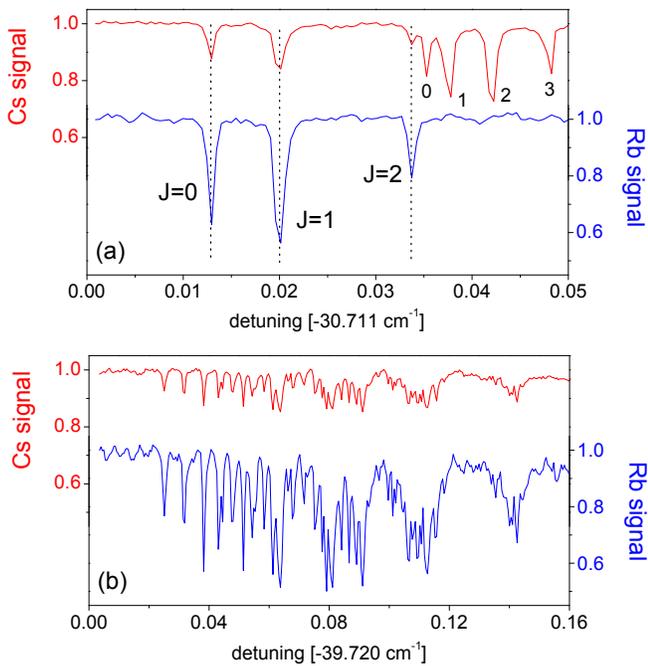}
\caption{(Color online) RbCs$^*$ PA signals. The PA detuning is specified relative to the 6S$_{1/2}$, $F=3\to$ 6P$_{1/2}$, $F^\prime=3$ transition of Cs at 11178.4172 cm$^{-1}$. The upper traces (left axes) show the Cs trap population and the lower (right axes) the Rb trap population. (a) The features marked with dashed lines arise from the $J=0,1,2$ rotational components ($J$ is the quantum number associated with the total molecular angular momentum, except nuclear spin) of a RbCs$^*$ $0^-$ vibrational level with an outer turning point at 13.2\AA. Another rotational series also appears only in the Cs trace, associated with a $0_u^+$ level of Cs$_2^*$ having an outer turning point at 24.8\AA. Of the two, the RbCs$^*$ state has a much larger rotational splitting due to its shorter-range character. (b) Resolved hyperfine-rotational substructure of a RbCs$^*$ level with $\Omega\neq 0$.}
\label{fig2}
\end{figure}

In order to detect PA-induced loss, the fluorescence
rate of each MOT was monitored with a photodiode, as the frequency
of the PA laser was scanned. An example of the observed signals is
shown in Fig. \ref{fig2}, where the features common to both Rb and
Cs traces can be identified as RbCs$^*$ states. From the depletion
of the Rb trap (up to 70\%) we estimate resonant heteronuclear PA
rates as large as $\sim 1.5\times 10^8$ s$^{-1}$, close to the
expected maximum rate for our parameters determined by probability
conservation in a binary scattering process (the so-called
``unitarity limit") \cite{bohnPA,epaps}. The loss rates are indeed
observed to saturate as a function of intensity, at values
typically $> 10^6$ W/m$^2$, significantly higher than in
homonuclear experiments \cite{molsum,Li6Li7}. These values are
also consistent with our estimates based on a WKB approximation
for the ground-state wavefunction at short range
\cite{epaps,wangstwa,bohnPA}.

We have observed RbCs$^*$ levels over the range $\Delta\sim 10 \to
100$ cm$^{-1}$ \cite{epaps}, and we find vibrational progressions
corresponding to the expected $\Omega = 0^+, 0^-, 1$, and $2$
potentials dissociating to both the 6P$_{1/2}$ and 6P$_{3/2}$
limits, as shown in Fig \ref{fig1}(b). The $\Omega=0^\pm$ levels
have no hyperfine splitting in leading order, and are thus
identified by their clean rotational structure [Fig
\ref{fig2}(a)]. The $\Omega = 1,2$ features display a complex
hyperfine-rotational structure, [Fig. \ref{fig2}(b)], the analysis
of which is in progress.

\begin{figure}
\includegraphics[width=3.375in]{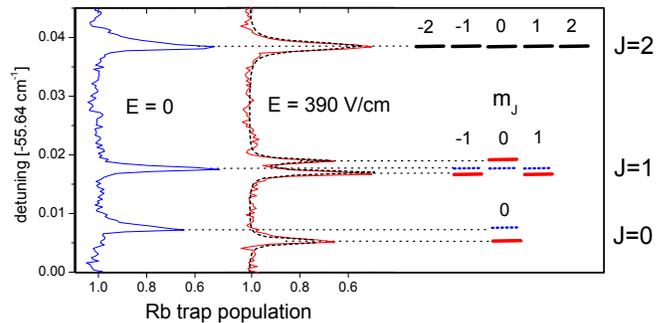}
\caption{(Color online) Stark effect in a RbCs$^*$ $\Omega=0^+$
state.  The dotted line is a fit of the $E=390$ V/cm spectrum,
based on the expected Stark effect for a diatomic rigid rotor
(illustrated schematically on the right) with $\mu_e =1.3$ Debye.}
\label{fig3}
\end{figure}

We have also demonstrated the polar nature of the observed
RbCs$^*$ states by applying electric fields [Fig \ref{fig3}]. The
observed Stark effect agrees in form with that expected for a
polar, diatomic rigid rotor, as illustrated by the fit in Fig
\ref{fig3}. From this fit we extract an electric dipole moment for
this level of $\mu_e=1.3\pm 0.1$ Debye.

\begin{figure}
\includegraphics[width=3.375in]{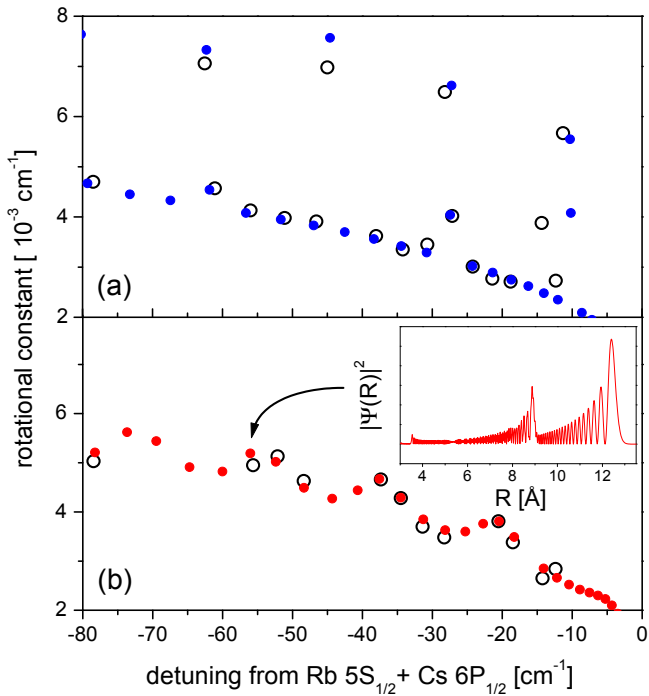}
\caption{(Color online) Comparison between observed $\Omega=0^\pm$ levels (open circles) and those calculated from  fitted RbCs$^*$ potentials (solid circles). Each point corresponds to a rotational series (e.g. Fig. \ref{fig2}(a)); the horizontal axis gives the energy $E_{v,0}$ of its $J=0$ component, and the vertical axis its rotational constant $B_v$, defined by: $E_{v,J}-E_{v,0} = B_vJ(J+1)$. (a) $0^-$ levels; (b) $0^+$ levels. Inset: $|\Psi_{b,e}(R)|^2$ for the $0^+$ level at $\Delta=-55.63$ cm$^{-1}$, showing its strongly mixed character.}
\label{fig4}
\end{figure}

We have analyzed our observations of the $\Omega =0^\pm$ levels by
fitting them to a model of the RbCs$^*$ potentials, based on
\textit{ab initio} calculations \cite{RbCspot} and previous
spectroscopic data \cite{tom,epaps}. A thorough discussion of this
analysis will be presented elsewhere. Fig. \ref{fig4} shows a
comparison between the observed $\Omega=0^\pm$ levels, and the best
fit from our model. In
the $0^-$ case [Fig. \ref{fig4}(a)], two distinct vibrational
series are evident, with spacings on the order of 3 cm$^{-1}$ and
15 cm$^{-1}$. These are associated with the two different $0^-$
potentials dissociating to the 6P$_{1/2}$ and 6P$_{3/2}$ atomic
limits [see Fig. \ref{fig1}(b)]. The latter are bound by more
than 550 cm$^{-1}$, and have outer turning points as small as
9\AA\; at our detunings; consequently, these features would be
difficult to see without our high PA intensity, as their
free-bound FCFs are relatively small. Also evident in the figure
is a coupling between these two vibrational series, causing a
perturbation in the rotational constant when a near-degeneracy
occurs.  For the $0^+$ levels [Fig. \ref{fig4}(b)], the coupling
is so much stronger that almost no trace remains of the
``unperturbed" vibrational structure, and all levels have a
strongly mixed character, as illustrated by the wavefunction in
the inset. This type of coupling has been discussed theoretically
\cite{molsum}, and was previously observed in $0_u^+$ levels of
Cs$_2^*$ \cite{Cs2}, though it is a much weaker effect in that system. As a final
note, the values we extract from our analysis for the long-range
dispersion coefficient $C_6$ of the RbCs$^*$ A$^1\Sigma_0^+$,
b$^3\Pi_0$, and (2)$^3\Sigma_0^+$ states [see Fig. \ref{fig1}(a)]
all agree with \textit{ab initio} calculations \cite{marsad} to
within a few percent \cite{C6,epaps}.

Using our model of the RbCs$^*$ potentials, we can make
quantitative estimates of the rate at which ground-state molecules
are formed via spontaneous decay in our experiments. For example,
taking our observed PA rate of $\sim 1.5\times 10^8$ s$^{-1}$ on
the $\Omega=0^+$, $J=1$ resonance at $\Delta=-55.63$ cm$^{-1}$, we
predict that X$^1\Sigma^+$ state molecules will be formed in
vibrational levels near $v = 62$, bound by almost 1300 cm$^{-1}$,
at a rate of $\sim 5\times 10^5$ s$^{-1}$ per level \cite{stim}. In contrast
to the homonuclear case, extremely deeply bound molecules are
formed at large rates via decay at a spatially coincident
\textit{inner} vibrational turning point of the ground and excited
levels \cite{omega1}. This arises from the
shorter-range character of the asymptotic potential as well as the 
strong $0^+$ channel coupling discussed above, which
together can produce a large probability density at the inner
turning point of the A$^1\Sigma^+$ excited state potential. The
resulting deeply bound ground state molecules in $J=0,2$
rovibrational levels could be easily transferred with high
efficiency to $v=0, J=0$ using a single two-photon Raman
transition; for example, we calculate that the two FCFs for such a
transition, via an intermediate state at $\Delta\sim 3900$
cm$^{-1}$, are both $\geq 10^{-2}$. These transitions could easily
be saturated using standard pulsed lasers.

In summary, we have demonstrated photoassociation into polar
RbCs$^*$ molecular levels in a laser-cooled, dense mixture of
$^{85}$Rb and $^{133}$Cs atoms, at large rates up to $\sim
1.5\times 10^8$ s$^{-1}$. Analysis of our spectra indicates
molecule formation rates via spontaneous decay into deeply bound
rovibrational levels of the ground X$^1\Sigma^+$ state as high as
$5\times 10^5$ s$^{-1}$ per level, due to the inherently
short-range character of the RbCs$^*$ levels we excite. We plan to
detect these ground state molecules using resonance-enhanced
two-photon ionization \cite{Cs2,K2,Rb2,Na2} and to trap them in an
optical dipole trap. A two-photon Raman transition from one of the
well-populated levels to the rovibrational ground state should
allow us to produce a large sample of stable, ultracold polar
molecules.

We thank O. Dulieu for information on Cs$_2^*$ potentials, and M. Pichler and W. Stwalley for the use of the former's Ph.D. thesis. We acknowledge support from NSF grant EIA-0081332, and the David and Lucile Packard Foundation. T.B. acknowledges funding from the U.S. Office of Naval Research.


\begin{thebibliography}{}
\bibitem{stim} We neglect stimulated processes (such as photodissociation) which may also deplete the excited levels, since we observe no PA-induced broadening at the few MHz level.
\bibitem{csint} The lower atom number and density and higher temperature of the Cs MOT resulted from its higher trapping laser intensity; this was used to keep its intrinsic losses high, to reduce its depletion by the many Cs$_2^*$ resonances.
\end{thebibliography}
\end{document}